# HARMONI at ELT: Developing a Modular Review Process using TRIZ

Amelia Calderhead*, Anna MacIver, David Isherwood, Andy Born, James Stevenson

*UK Astronomy Technology Ctr . on behalf of the HARMONI consortium*

## ABSTRACT

HARMONI is the first light, adaptive optics assisted, near-IR integral field spectrograph for the ELT. It covers a spectral range from 800nm to 2450nm with resolving powers from 3000 to 7000 and spatial sampling of 25mas and 6mas. It can operate in two adaptive optics modes - SCAO (including a high contrast capability) and MCAO. The project is resuming its final design phase after a rescope design phase in 2025.

The HARMONI team at UKATC have been pioneering the use of TRIZ by managing monthly TRIZ workshops where colleagues from all disciplines look to tackle significant project problems that arise. This paper will be on the TRIZ case study of investigating how design changes of HARMONI can be reviewed effectively and efficiently. It draws on how the review process of Ground Based Astronomy projects can be improved to add quality to the successful creation of an instrument while reducing strain on designers and reviewer's time.

The existing process surrounding the traditional design reviews, based on rigid milestones such as Critical Design Review (CDR), has proven to be resource-intensive and heavily document-driven. These reviews often suffered from late detection of technical issues, resulting in costly redesigns and rework. To address these issues, the case study utilised TRIZ techniques including; thinking in time and scale technique: the 40 inventive principles to tackle design contradiction: function analysis and size, time, cost analysis.

The solution developed considered a modular, continuous review process focusing on specific subsystems or technical topics as they mature. The final proposal allows milestones to remain in place for procurement and contractual alignment, but they will serve as deadlines for consolidating progress rather than as bottlenecks. This approach encourages a live review culture, where stakeholders and expert reviewers are engaged continuously through informal check-ins rather than infrequent formal meetings.

*Email: amelia.calderhead@stfc.ac.uk

# 1. INTRODUCTION

The traditional approach to design reviews is based on rigid milestones such as Preliminary Design Review (PDR) and Critical Design Review (CDR) and has proven to be resource-intensive and heavily document-driven.

An example of the challenges faced is the HARMONI Rescope Process which was initiated in 2024 (Neichel, et al., 2026). This was identified as an opportunity for improvement in not only design, but how reviewing the design was approached. Figure 1-1 shows the rescope schedule for HARMONI.

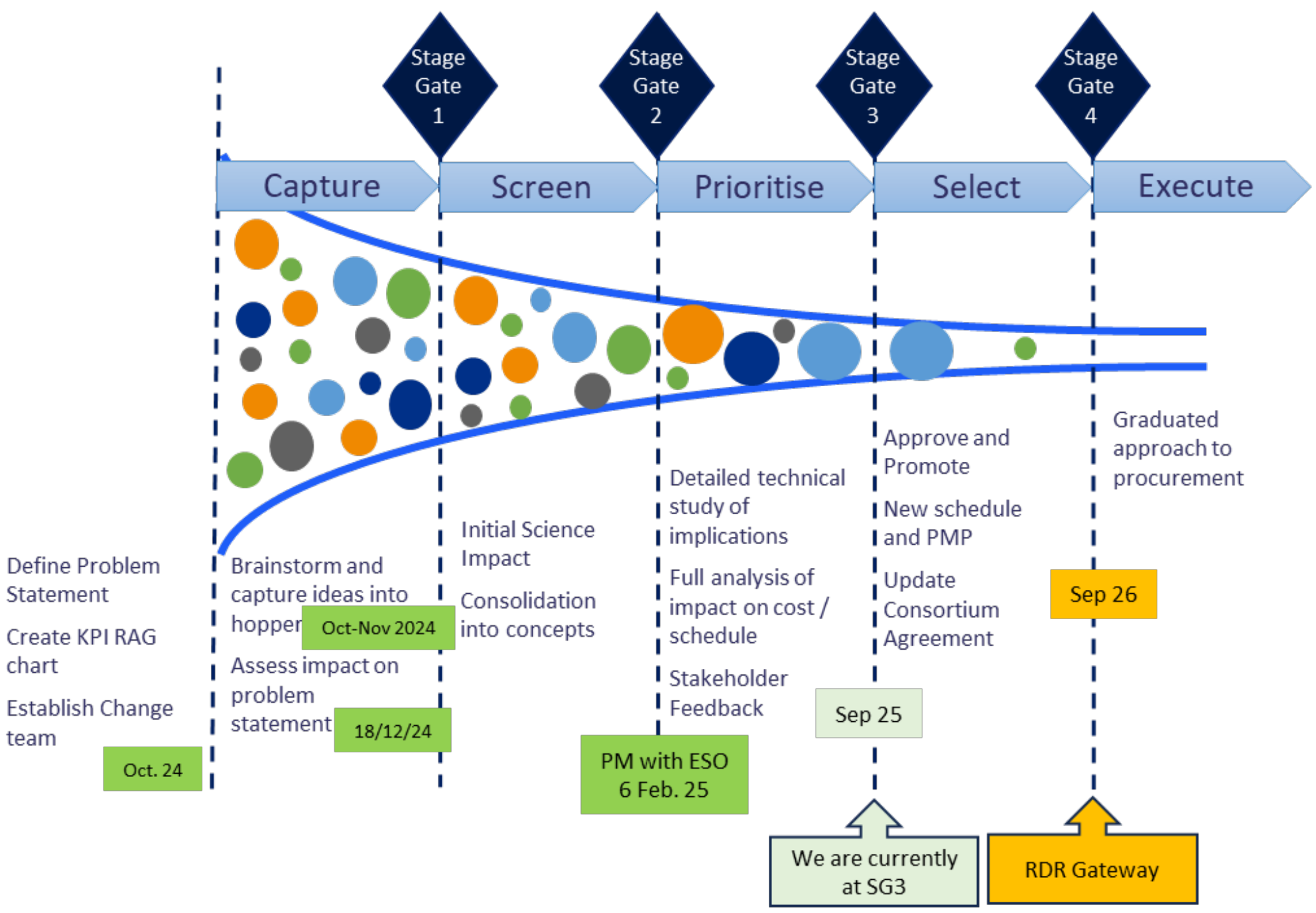

**Figure 1-1 HARMONI Rescope Timeline (Chittick, 2025)**

During the rescope HARMONI has undergone a significant redesign which invalidates some of the previously conducted PDR / CDR reviews. Over just 2 years the project aims to capture, screen, prioritise, select and execute a new version of the instrument leading up to the Rescope Design Review (RDR) in October 2026.

The objectives of RDR are to:

- Provide preliminary design for sub-systems under-going major redesign
- Demonstrate Re-scoped HARMONI design compliancy

It was identified the consortium has limited resource to produce and review large amounts of paperwork which is common with the design review process but even more significant when the whole instrument is being reviewed. It was decided to use theory of inventive problem solving (TRIZ) approaches to investigate and find solutions to the challenge as a part of the UKATC TRIZ working group.

## 2. TRIZ PROBLEM

The first step was to define and understand the TRIZ problem to be solved.

The TRIZ Problem

*How to* ***efficiently and effectively review and validate design changes*** *made since the original PDR, without redoing the entire PDR review*

This forced us to review some issues with previous design reviews and apply TRIZ tools for problem solving.

The TRIZ team applied the TRIZ examples to look for possible improvements. This started by assessing if there was anything wrong with previous reviews throughout a wide range of astronomy projects and experience. It was found that previous reviews were resource-intensive and heavily document-driven. Previous reviews generated thousands of Review Item Discrepancies (RIXes), most of which focused on documentation rather than technical compliance. These reviews often suffered from late detection of technical issues, resulting in costly redesigns and rework. Political pressure to pass reviews and limited resources further reduced their effectiveness. The main challenge is to validate design changes efficiently, checking for key technical issues and adding quality to the instrument.

## 3. TRIZ METHODOLOGY

TRIZ was primarily the work of the engineer Genrich Altshuller and provides a variety of ways to find solutions to a wide range of problems (Gadd, 2011). It was chosen to utilise the following TRIZ techniques; thinking in time & scale technique: the 40 inventive principles to tackle design contradiction: function analysis and size, time, cost analysis.

Thinking in Time and Scale (9 Box Exercise)

This considers the problem in the present, past and future as well as at a system level, supersystem level and sub system level (Gadd, 2011). The conclusions of the exercise are seen in

Table 3-1.

**Table 3-1 9 box exercise**

| **Level** | **Past (PDRs in the past)** | **Present (Delta PDR)** | **Future (Ideal State)** |
| --- | --- | --- | --- |
| **Super System (SS)** | *Processes very strictly defined*<br>*RIXes captured in spreadsheet*<br>*Debates about interface and requirements dominate review*<br>*Last-minute redesign* | *Political Pressure to complete review and pass*<br>*Financial Risk to not passing*<br>*RIXes in JIRA*<br>*25–30 slide presentations*<br>*Expectation on certain analysis and testing*<br>*Low resource availability* | *Reduced political pressure to pass*<br>*No debate about requirements or interfaces*<br>*Clear expectations flowed down*<br>*Clear technical feedback*<br>*Experts aware and on side*<br>*Less Formal*<br>*The review is a deadline rather than formal review* |
| **System (S)** | *1000+ RIXES*<br>*100–200 drawings*<br>*CDR*<br>*2 one-week meetings* | *100 RIXES open*<br>*CDR – 1 per sub-system*<br>*Document based review*<br>*Requirement flow down not clear* | *Less documents/Not just documents*<br>*Compliance against requirements clear*<br>*No questions*<br>*Less RIXES – only important issues* |
| **Sub-System (sS)** | *Calibration plan and instrument overview resulted in most questions*<br>*Instrument description*<br>*RIXES about document issues / not technical*<br>*Not good conclusions* | *Detailed DARS + documents*<br>*Sub-systems NOT taking responsibility for deciding technical content*<br>*Demand for templates* | *Clear Requirements*<br>*Manageable number of RIXES* |

A need for a more informal review with a clearer feedback and RIXes process was identified as a necessary approach.

Technical Contradiction

The 40 inventive principles can be used to solve contradictions. The process first identifies two contradicting principles then a range of inventive principles can be identified (Gadd, 2011) as seen in Table 3-2. The solutions were then discussed around the key inventive principles that were found in Table 3-3.

**Table 3-2 Identification of inventive principles**

| **Improve** | **Without Worsening** | **TRIZ Inventive Principles** |
|---|---|---|
| 24. Loss of information | 25. Loss of time | 10. Prior Action, 28. Replace System, 35. Parameter Change |
| 37. Difficulty of detecting | 25. Loss of time | 3. Local Quality, 23. Feedback, 24. Intermediary |
| 35. Adaptability or versatility | 25. Loss of time | 15. Dynamics, 6. Universality, 35. Parameter Change |
| 35. Adaptability or versatility | 33. Convenience of use | 32. Colour Change, 13. The Other Way Round |
| 38. Extent of automation | 24. Loss of information | 25. Self-Service, 11. Cushion in Advance, 10. Prior Action |
| 39. Productivity | 24. Loss of information | 2. Taking Out, 9. Prior Counteraction, 10. Prior Action |
| 33. Convenience of use | 24. Loss of information | 33. Homogeneity |

**Table 3-3 Inventive principle solutions**

| **TRIZ Inventive Principles** | **Apply to this setting** |
|---|---|
| 35. Parameter Change<br>Modify system parameters to optimize performance. | *Change inputs from formal document into technical notes, continually updated*<br>*Change documents to model based approach* |
| 32. Colour Change | *Highlight aspects which have changed*<br>*More transparency into ongoing work* |
| 28. Replace Mechanical System | *Substituting manual, rigid, or mechanical processes with more flexible, digital, or automated systems* |
| 10. Prior Action | *Pre assign reviewers* |

It was identified that a transparent model based approach where aspects which have changes are highlighted, and reviewers are pre assigned were needed for the overall approach.

Size, Time, Cost

This exercise considers there are no limitations to size, time and costs to generate solutions using infinite time and cost scenarios. This is to break psychological inertia and not ignore different possibilities which seem out of reach (Gadd, 2011). Table 3-4 summarises what was discussed.

**Table 3-4 Size, Time, Cost Analysis**

| No Budget | Intermediate Solution | Infinite Budget |
|---|---|---|
| *Sharing resources with other projects* | *Prototyping: condensing docs, specifying which docs to review, wider reviews, customer involvement* | *No PDR until built & functioning*<br>*Identify issues via test*<br>*Infinite resources to review all*<br>*Special experts available 24/7 ( helpline)* |
| *Remote reviews*<br>*Provide open access to everything we've done rather than generating specific documents* | *AI-generated documents from known inputs & requirements*<br>*Modular design review* | *Overseers/reviewers for each sub-system + system level*<br>*Automated reviews of CAD/hardware*<br>*AI involvement*<br>*No documents or descriptions* |
| *1 design review in the whole project*<br>*AI reviews everything – no human resource* | *Documentation ongoing*<br>*No formal design review* | *(Blank)* |

The intermediate solutions identified were;

1. Prototyping to show compliance
2. Condensing documents to have a compliance based review
3. Specifying which documents to review
4. Wider reviews cross project
5. Customer involvement earlier in process
6. AI-generated documents from known inputs & requirements
7. Modular design reviews
8. Documentation ongoing with no formal design review

Solutions 3, 5, 6, 7 and 8 were brought forward for the approach.

# 4. APPROACH

Realising that projects are always going to be resource constrained, to make a more effective review process a shift would be beneficial. To address the problems identified, the proposal was to shift from milestone-based reviews to a more agile **modular, continuous review process** which reduces review burden, and improves traceability. Milestones will remain in place for procurement and contractual alignment, but they will serve as deadlines for consolidating progress rather than as bottlenecks.

Modular reviews will focus on specific subsystems or technical topics as they mature, reducing the pressure to achieve perfection by a single date. This approach encourages a live review culture, where stakeholders and expert reviewers are engaged continuously through informal check-ins rather than infrequent formal meetings. Documentation will be streamlined by capturing outcomes in technical notes and presentations, with the potential use of AI tools to consolidate these into formal milestone documents. The goal was to still meet formal and procurement-related requirements as the terminology is constrained, however these are seen as deadlines rather than reviews. The key principles are described in Table 4-1.

**Table 4-1 Principle Descriptions**

| Principle | Description |
|---|---|
| Milestone Reviews as Checkpoints, Not Bottlenecks | PDR, CDR, etc. are retained for political and procurement alignment. These are treated as **deadlines for consolidation**, not the only review moments. |
| Modular Reviews for Incremental Progress | Conduct **targeted reviews** of subsystems, models, or analyses as they mature. Each review is scoped to a specific technical area or design increment. Reduces pressure to "perfect" everything by a single date. |
| Live Review Culture | Encourage **ongoing engagement** with reviewers and stakeholders. Replace infrequent formal meetings with **regular, informal check-ins**. Keep customer and expert reviewers informed continuously. |
| Documentation Strategy | Use presentations and technical notes to capture modular review outcomes. Explore AI tools to consolidate these into formal milestone documentation. Final milestone documents become compliance based documents, with high-level summaries against requirements. |
| Review Burden Reduction | Spread review effort over time. Avoid duplication and rework by reviewing components when ready. Focus milestone reviews on integration and requirement closure, not re-review. |
| Integration with Project Tools | Link modular reviews to JIRA tickets or epics. Define cadence (e.g. one review every 6 months per subsystem). Use JIRA to track review status and inclusion in milestone documentation. |
| Panel-Based Review Model | Small, competent panels review specific aspects. Reviewers are selected based on expertise, not availability. RIXes (Review Item Discrepancies) are raised only for critical actions, accumulated over time. Each modular review is linked to JIRA tickets and tech notes if necessary. |
| Failure is an option | Due to small modular reviews, failing modular reviews becomes standard practice and issues can be flagged timely. |

Function Analysis

Functional analysis maps systems in order to identify problems with system functions by breaking them down into subject → action → object (Gadd, 2011). The action can either be harmful (red), normal (green) or insufficient (dashed).

A function analysis map was created once we had an approach proposal as seen in Figure 4-1. This was to assess and suggest appropriate responsibilities and relationships between job roles, inputs, and outputs. This identified where there were weaknesses in our proposed modular review.

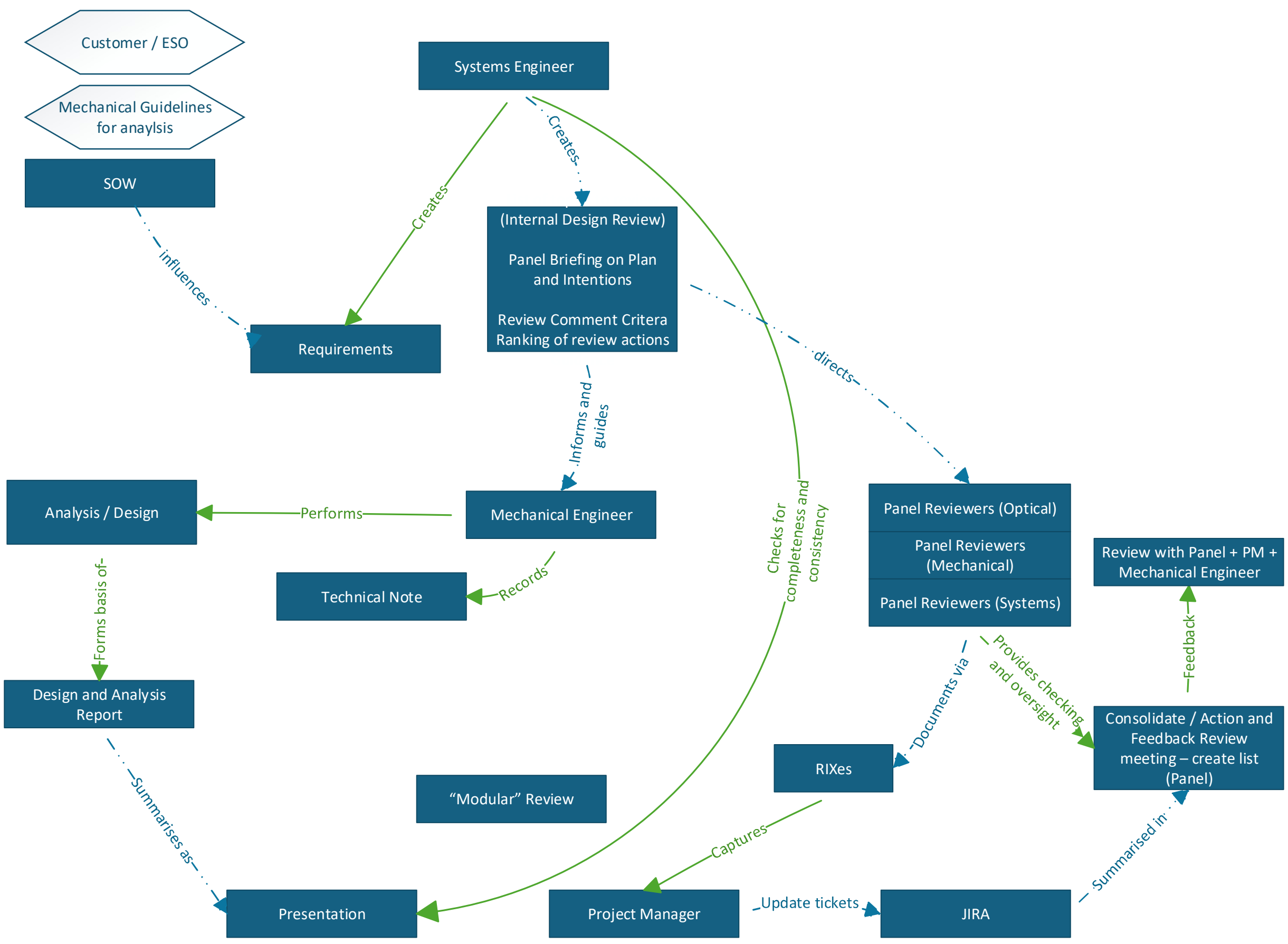

**Figure 4-1 Function Analysis Map**

## Potential Issues

The function analysis map allowed potential issues to be identified. An increase in review panels needed means there is a burden of review panels to be available more frequently creating a greater insufficiency. This impact can be reduced and done more sufficiently by creating appropriate scheduling of reviews and reviewers time. I.e. an optical engineer probably does not have to be present for most of a review of a cryostat design. However this also increases the burden of planning topics and reviews in advance so more resources may be needed to manage this effectively. There may also be more need for more minute and action taking than previously. These could be delegated to a project assistant adding a new subject to the map. The range of documentation reviewed may be greater over time but will be mitigated by more continuous documentation throughout the design process not just before a review is planned. There is also a risk this approach is not bought in by stakeholders therefore we aim to apply it to internal reviews as a trial.

## Trial Plan

The trial will apply modular reviews to selected subsystems or projects. Reviews will be scheduled at regular intervals. Each review will be linked to JIRA tickets and documented through technical notes or presentations. This approach will allow the team to evaluate reviewer engagement, documentation quality, time and resource savings, and the effectiveness of risk identification.

## Key Process

1. Identify sub-systems / modules for review
2. Identify internal review panel
3. Create mechanical, optical & systems checklists based upon project requirements and maturity level
4. Run modular review panels

Final Proposal

To differentiate between reviews, the following terminology was introduced:

**RDR** - Rescope Design Review (Run by European Southern Observatory ESO)

**RaDAR** - Rescope Design Assessment Review (Run internally)

The RaDAR is an internally managed process. The proposed approach advocates for a more agile, modular review structure with smaller expert panels, continuous oversight, and reduced documentation burden. Subsystem reviews (RaDARs) will be managed by the system, however ESO are invited to participate, similar to the CDR process. The review will be tailored for each subsystem based on the extent of changes since the PDR.

An example of reviews done in the past in the IFS sub systems are seen in Table 4-2.

**Table 4-2 Previous Design Reviews on the IFS**

| | | **Reviews done in the past** | | |
|---|---|---|---|---|
| | Level of change | PDR | CDR | FDR |
| **System** | | | | |
| IFS | Moderate Change | x | | |
| **Sub-systems** | | | | |
| IPO | Moderate Change | x | x | |
| IFU | No Change | x | x | |
| ISP | Major Change | x | x | |
| ICR | Moderate Change | x | x | |
| ISDM | Minor | x | x | |
| PCAM | NEW system - no design | | | |

Subsystem RaDARs will be split into modules covering optics, mechanical design, and systems compliance. These inputs will feed into the system-level RDR, which will assess overall performance, integration, and verification planning.

To maximise efficiency, the panel should be kept small. Each of the panels will have:

- Sub-System presenter
- IFS Systems Engineer
- Lead discipline engineer
- 1-2 Additional domain specialists (can include ESO follow up team)

The content of these reviews is given below.

### Sub-System Mechanical RaDAR Content

The review will be called to review the key mechanical analyses.

The review will include:

- Mechanical Checklist Items identified
- Review of CAD
- Compliance to Space Envelope - Overall layout - mass, size (fits within allocated space envelope)
- Compliance to Mechanical and Thermal ICD

An example checklist for a IFS sub system in HARMONI is below.

**Table 4-3 Mechanical Checklist Example**

| Item | Guidance |
|---|---|
| *Earthquake structural analysis* | *Preliminary analysis of earthquake structural analysis as defined in RD3 section 4.2 to demonstrate no show stoppers* |
| *Modal Analysis* | *The first resonances of the subsystem will be calculated as per RD3 section 4.9* |
| *Thermal & Structural Analysis* | *The thermal analysis will be conducted in line with RD3 sections 4.10, 4.11 and 4.12*<br>*Critical is optical face distortion when mounted & cold* |
| *Mounting key calculations* | *If any updates to mounting to cold structure, review calculations - can occur of source material in RADAR review or before (see section 5)*<br>*Design of Sub-system position adjustment/structural mounts, including review of key calculations (structural fastener size)* |
| *Materials* | *List of materials shared with LDE*<br>*Key here is anything not aluminium, (e.g. stainless, copper, G10, PCB's, or cables, how interfaces between dissimilar materials are handled. (e.g. optics mounts))* |
| *Mechanical ICD*<br>*Thermal ICD*<br>*Space Envelope ICD* | *Input into defining new ICDs & compliance* |

## Sub-System Optical RaDAR Content

The review will be called to review the key optical analyses.

The review will include:

1. Optical Checklist Items identified
2. Review of Zemax Model (submit in advance)
3. Optical ICD

Table 4-4 shows an example optical checklist for HARMONI IFS sub systems.

**Table 4-4 Optical Checklist example**

| **Item** | **Guidance** |
|---|---|
| *Summary of the optical design changes & first order parameters* | *AD2 Section 3.1.2* |
| *Nominal Subsystems performance description* | *AD2 Section 3.1.3* |
| *Tolerance Analysis for as built performance* | *AD2 Section 3.1.6* |
| *Delivery of the optical models for input into E2E Model* | *AD2 Section 3.5* |
| *Optical Alignment Plan* | *AD2 Section 3.1.5* |
| *Assessment against key requirements (KPIs)* | *See below* |

## Sub-System Systems RaDAR Content

The review will be called once all the documentation is completed. This will be the final check of the following:

1. Check that all paperwork has been updated to the necessary level needed to feed in to IFS documents
2. Review of Verification Matrix
3. Review of sub-system compliance matrix
4. Review of key technical risks against Verification testing status

### JIRA RIX

It was also proposed how to add RIXes to Jira as part of the RaDAR review.

RIXes should not be created until the review board have assessed all comments and agreed upon a set of tasks to be completed by the engineers.

A RIX is only a RIX if there is a clear deliverable. If questions can be answered prior to a RIX being created, this is preferable. A RIX is not an opinion, it is a constructive comment which can lead to improvement in system and correction of errors.

## Proposed RIX format

**Raised RIX:** Here is a brief introduction with context, the question or observation should be raised and a general task described.

**Reference:** The section of DAR or tech note should be specified.

**Stakeholders:** Including the original RIX raiser and members of the panel that are concerned with the results of this RIX. This should be identified by the panel. This is who the answer should be circulated with or discussions should include.

**Task description**:

Here a breakdown of the tasks anticipated to resolve this should be set out.

This is to save time and pressure on the engineer, especially as the RIX raiser may have in mind what needs to be done here.

Suggested tasks could include; adding additional information to the report, repeating analysis' with or without different conditions, updating CAD models or drawings etc.

**Deliverable**: Output to be done by the original engineer or counterpart.

**Definition of complete**:

How the output is considered complete.

The next steps to move to checking.

**Definition of checking**:

Who will check and confirm RIX is closed.

This could include who to go to if RIX raiser is no longer in formal role.

## Example RIX

**Raised RIX:** The values for the first column of the model are very high and there could be an error. Complete a reanalysis of the model checking the inputs.

**Reference:** System Model Technical Note, Section 3.5.2, Table 3-4

**Stakeholders:** Anna MacIver, Amelia Calderhead, Andy Born

**Task description**:

Check inputs with original requirements and CBEs

Re analyse model

Add results to tech note

**Deliverable**: Updated results to be sent to stakeholders

**Definition of complete**:

Model reanalysed

Results sent to stakeholders

**Definition of checking**:

Stakeholders review results

RIX raiser confirms results

# 5. CURRENT STATUS & CONCLUSIONS

The progress of planning RaDARs for the IFS sub systems has progressed well. All sub systems have decided the dates of their review periods, commencing the end of June 2026. Key reviewers have been identified and their availability has been discussed and confirmed. The process so far has allowed for sub systems to be realistic about how they will prepare their documentation for RDR and set achievable but stretching goals that should allow for a successful review in the future.

The work of the TRIZ working group has been beneficial and this specific case study has allowed the team to reflect on the importance of reporting and reviewing successful astronomical design work. The use of multiple TRIZ tools such as thinking in time & scale, the 40 inventive principles, function analysis and size, time, cost analysis showed the variety of ways problem solving solutions can be generated & developed to optimise the HARMONI rescope review and how TRIZ is not just an application for physical engineering problems. The solution developed for RaDARs will use a modular, continuous review process focusing on specific topics as they mature while encouraging informal check-ins. The final review RIXes will be consolidated and added to Jira where these can then be managed as well defined pieces of work. Overall the trial will be completed by the end of September 2026 and a review of the impact this has made can then be facilitated.